# Modulation instability in nonlinear positive-negative index waveguide array

**A A Dovgiy[1] and A I Maimistov[1,2]**

[1] *National Research Nuclear University MEPHI (Moscow Engineering Physics Institute), 115409, Moscow, Russian Federation*

[2] *Moscow Institute for Physics and Technology, 141700, Dolgoprudny, Moscow region, Russian Federation*

E-mail: dovgiyalexandr@gmail.com, aimaimistov@gmail.com

Modulation instability in a nonlinear optical waveguide array with alternating positive and negative refractive indices is investigated analytically. Particular solutions of a system of coupled nonlinear equations are found. These solutions describe propagation of the continuum backward and forward waves in longitudinal direction of considered system. Instability of these waves is investigated with a linear stability analysis. A gain spectrum depending on different parameters of considered system: electromagnetic field power and ratio of the backward to forward wave amplitude, is studied analytically and numerically. It's shown that in case of Bragg resonance the continuum waves are stable. If nonlinearity assists at one kind of waveguides only and another one is linear one then the modulation instability doesn't occur. Out of Bragg resonance in defocusing waveguides the instability of the continuum waves is reduced.

## 1. Introduction

Modulation instability (MI) is a nonlinear wave phenomenon that occurs in many nonlinear systems, for example, in neutrino – antineutrino interactions [1], atomic vapors [2], pulsar plasma [3], laser – plasma interaction [4], easy – axis antiferromagnetic chains [5], Bose – Einstein condensates [6, 7], hydrogen bonded crystals and deoxyribonucleic acid [8]. In case of nonlinear optics, MI has been observed in fibers [9], liquid crystals [10], photorefractive crystals [11], nonlinear cavities [12], quadratic and nonlinear Kerr media [13, 14]. MI occurs when a perturbed continuous wave experiences an instability that leads to an exponential growth of its amplitude or phase with the perturbations due to the interplay between nonlinear interaction and dispersive/diffraction effects. In nonlinear optical fibers this instability leads to beam break-up in either space or time, which for one's part leads to formation of the solitary waves (solitons). Therefore, the MI process is known to be the first step towards energy localization [15, 16].

In recent years, nonlinear discrete structures have been the subject of considerable research; (see, e.g., [17, 18] and references therein). These structures are encountered in various fields of physics. Wave propagation in optical nonlinear discrete systems exhibits a wealth of new features that have no analogues in continuous systems including the modified discrete dynamics of solitons, which have attracted much attention because of there potential for all-optical switching [19]. In periodic discrete structures, the extended states are Floquet – Bloch (FB) modes. In case of arrays of nearest-neighbor coupled nonlinear optical waveguides, the linear FB modes form a transmission spectrum consisting of allowed bands, separated by gaps where light propagation is forbidden. The transverse wave vector determines the type of discrete diffraction, thus it can be either normal or anomalous in each band, moreover, in a limited region nearly diffraction-free propagation becomes possible, too [19]. In nonlinear discrete structures the nonlinear interaction may also cause MI [19, 20], known as discrete MI. The type of discrete diffraction has a strong influence on the stability of FB modes, thus they can be stable in case of anomalous diffraction and self-focusing nonlinearity or in case of normal diffraction and self-defocusing nonlinearity, and unstable otherwise [20]. This is a unique feature of discrete systems and it has experimental acknowledgements. In optical case, it was found that MI appears in AlGaAs waveguide array with self-focusing Kerr nonlinearity and in case of normal diffraction propagating regime [21], on the other hand, in the anomalous diffraction regime MI was found to be totally absent even at very high power levels. Another experimental investigation has demonstrated that in $LiNbO_3$ waveguide array with self-defocusing saturable nonlinearity MI appears in anomalous diffraction regions of photonic lattice (at the edges of first Brillouin zone) [22], and, otherwise, MI does not develop in regions of normal diffraction [23].

In lattice systems, different discrete models exhibit different peculiarities in MI process leading to different conditions for the existence of localized modes, known as discrete solitons [24, 25]. The simplest model of discrete nonlinear optical lattice is a nonlinear waveguide array (NOWA) of identical waveguides with equal spaces between them. In [26] MI in such NOWA with self-focusing Kerr nonlinearity is investigated. It is shown, that transverse wave vector influences essentially on the stability of moving plane waves and solitary waves. The transverse wave vector determines physically the kind of discrete diffraction, positive or negative one, taking place in the array due to the transverse coupling. Thus, it is shown that there are two different scenarios for the evolution of modulationally unstable solitary waves in accordance with the kind of discrete diffraction.

In [27] authors have considered MI in nonlinear oppositely-directional coupler (NODC) with a negative index metamaterial (NIM) waveguide. NODC allows considering the interaction between forward- and backward-propagating coupled waves, in contrast to conventional one, where the interaction between forward-propagating coupled waves takes place only. It is shown that MI in NODC is significantly influenced by the ratio of the forward- to backward-propagating wave's power and the nonlinear parameters. MI occurs only for finite values of power in waveguides when the NIM waveguide is self-defocusing and another one is self-focusing and the ratio is negative. With increasing the power in waveguides MI is suppressed, while in the conventional directional coupler MI is amplified. A presence of the NIM waveguide provides more ways to manipulate MI and soliton formation in couplers with NIM channels.

In [28, 29] the NOWA with alternating positive index material (PIM) and NIM waveguides was considered. It was shown that, due to a presence of a band gap in a linear spectrum in such NOWA, new kinds of discrete solitons exist [28]. In addition, it was found that in such NOWA a slit soliton can propagate [29]. This solitary wave is a new example of known gap solitons.

In this paper, MI in the NOWA with alternating PIM and NIM waveguides (figure 1) is investigated. We assume that waveguides are characterized by the Kerr nonlinearity. The instability of continuum waves propagating in the longitudinal direction of considered NOWA is investigated with the linear stability analysis. A gain spectrum depending on different parameters of NOWA is studied analytically and numerically.

Different variants of nonlinear response in the waveguides are studied. Also, the influence of the ratio of the backward to forward wave amplitude on the gain spectrum is investigated. It is shown that the MI process in the NOWA under consideration distinguishes from MI in NODC.

**2. Basic equations of the model**

The system of coupled equations describing the evolution of waves in the NOWA (figure 1, 2) involved in normalized variables has the following form [30]:

$$i(\partial_\zeta + \partial_\tau)a_n - \Delta a_n + K(b_{n-1} + b_n) + r_1 |a_n|^2 a_n = 0, \tag{1}$$

$$i(-\partial_\zeta + \partial_\tau)b_n - \Delta b_n + K(a_n + a_{n+1}) + r_2 |b_n|^2 b_n = 0, \tag{2}$$

where $a_n$ ($b_n$) is the normalized electric field envelope of the quasiharmonic wave in PIM (NIM) waveguide with number $n$; $r_1$ and $r_2$ are nonlinearity parameters of PIM and NIM waveguides, respectively; $K$ ($K > 0$) is a coupling coefficient between neighbor waveguides; $\Delta$ is a phase mismatch. Dissipations effects and the second-order group velocities dispersion are neglected. The details of the derivation of these equations can be found in [31]. There is the minus sign of spatial derivative in equation (2), unlike equation (1). This is because in NIM waveguide the backward wave propagates. While in PIM waveguide forward wave propagates. In this case phase velocity and the Pointing vector are directed in the same direction.

It should be noted that the metamaterials with the negative refractive index usually possess rather large losses. It is necessary to take into account the dissipation effects when light propagation in such metamaterials is investigated. However, in [32], the authors have demonstrated experimentally the existence of NIM in which optical waves propagate without losses. This brings high hopes that in nearest future it will be possible to manufacture optical waveguides of the metamaterial with negative index.

**3. Continuum waves and dispersion relation**

Particular solutions of equations (1) and (2) that describe the propagation of plane waves in the longitudinal direction of the NOWA under consideration have the following form:

$$a_n = a \exp(i\kappa\zeta + iqn - i\omega\tau) + c.c., \tag{3}$$

$$b_n = b \exp(i\kappa\zeta + iqn - i\omega\tau) + c.c., \tag{4}$$

where $q$ is a transverse wave number, $\kappa$ is a longitudinal wave number and $\omega$ is a deviation from the carrier frequency. Substitution of (3) and (4) to equations (1) and (2) with $r_1 = r_2 = 0$ allows obtaining the following dispersion relation for the linear waves:

$$\omega^{(+),(-)} = \Delta \pm \sqrt{\kappa^2 + 4K^2 \cos^2 \frac{q}{2}}.$$

It follows the spectrum of linear waves has the band gap (figure 3). The radiation of certain frequencies which belong to the band gap reflects from such NOWA. This feature is similar to the property of Bragg gratings, distributed mirror [33, 34] and NODC. However, when the transverse wave number takes the values: $q = \pi \pm 2\pi n$, $(n = 0,1,...)$, the spectrum has no band gap (figure 3). These values correspond to a Bragg resonance in the one dimensional lattice. When the transverse wave number takes any other values, the spectrum has a band gap of the width $\Delta\omega = 4K|\cos(q/2)|$. Maximum of the gap width is equal to $\Delta\omega_{max} = 4K$ when $q = 0$.

With increasing the wave's power in the waveguides it is necessary to take into account the nonlinear response which distorts the spectrum. Substitution of (3) and (4) to nonlinear equations (1) and (2) gives the following nonlinear dispersion relation:

$$\begin{cases} \omega = \Delta - K \dfrac{1+f^2}{f} \cos \dfrac{q}{2} - \dfrac{P}{2(1+f^2)} (r_1 + r_2 f^2), \\ \kappa = -K \dfrac{1-f^2}{f} \cos \dfrac{q}{2} + \dfrac{P}{2(1+f^2)} (r_1 - r_2 f^2), \end{cases}$$

where $P = a^2 + b^2$ is the total power in the couple of PIM and NIM waveguides and $f = b/a$ is the ratio of backward to forward wave amplitude. If $|f| < 1$ ($|b| < |a|$), then forward wave dominates in the array, and if $|f| > 1$ ($|b| > |a|$), then backward wave dominates in the array. A case $|f| = 1$ corresponds to absence of an energy transfer either in forward or in backward directions. In the lower power limit the linear spectrum and the nonlinear spectrum coincide with each other with good exactness (figure 3).

## 4. Linear stability analysis

In this section the instability of particular solutions (3) and (4) will be investigated with the standard linear stability analysis [16]. The amplitude perturbations of continuum waves (3) and (4) may be introduced as

$$a_n = (a + \tilde{a}_n)\exp(i\kappa\zeta + iqn - i\omega\tau), \qquad (5)$$

$$b_n = (b + \tilde{b}_n)\exp(i\kappa\zeta + iqn - i\omega\tau), \qquad (6)$$

where $\tilde{a}_n$ and $\tilde{b}_n$ are small quantities, i.e. $|\tilde{a}_n| \ll a$, $|\tilde{b}_n| \ll b$. Substitution of (5) and (6) to nonlinear equations (1) and (2) and linearization in $\tilde{a}_n$ and $\tilde{b}_n$ allows obtaining the following system of the linear differential equations:

$$\begin{cases} i(\partial_\zeta + \partial_\tau)\tilde{a}_n - f\mu \tilde{a}_n + K(\tilde{b}_{n-1} + \tilde{b}_n) + \rho_1(\tilde{a}_n + \tilde{a}_n^*) = 0, \\ i(-\partial_\zeta + \partial_\tau)\tilde{b}_n - \frac{\mu}{f}\tilde{b}_n + K(\tilde{a}_n + \tilde{a}_{n+1}) + \rho_2(\tilde{b}_n + \tilde{b}_n^*) = 0, \end{cases} \qquad (7)$$

where

$$\mu = 2K \cos\frac{q}{2}, \quad \rho_1 = r_1 \frac{P}{1+f^2}, \quad \rho_2 = r_2 \frac{Pf^2}{1+f^2}.$$

In order to solve the system of linear equations (7), we assume the plane wave ansatz which constitutes of both forward- and backward-propagating waves:

$$\tilde{a}_n = c_1 \exp(ik\zeta + iQn - i\nu\tau) + d_1 \exp(-ik\zeta - iQn + i\nu\tau), \qquad (8)$$

$$\tilde{b}_n = c_2 \exp(ik\zeta + iQn - i\nu\tau) + d_2 \exp(-ik\zeta - iQn + i\nu\tau), \qquad (9)$$

where $c_1, c_2, d_1, d_2$ are real constants, $Q$ is the perturbation transverse wave vector, $k$ is the propagation constant and $\nu$ is the perturbation frequency. Substitution of (8) and (9) to equations (7) leads to the system of four linear algebraic first order equations for $c_1, c_2, d_1, d_2$:

$$\begin{pmatrix} \nu - k - \mu f + \rho_1 & M & \rho_1 & 0 \\ M & \nu + k - \mu/f + \rho_2 & 0 & \rho_2 \\ \rho_1 & 0 & -\nu + k - \mu f + \rho_1 & M \\ 0 & \rho_2 & M & -\nu - k - \mu/f + \rho_2 \end{pmatrix} \begin{pmatrix} c_1 \\ c_2 \\ d_1 \\ d_2 \end{pmatrix} = 0,$$

where $M = 2K \cos Q/2$.

This system of algebraic equations has a nontrivial solution only when its determinant is equal to zero. It follows the algebraic fourth order equation for the perturbation frequency $\nu$ which determines the dispersion relation for the waves generated by small perturbations:

$$\nu^4 - a\nu^2 + b\nu + c = 0,$$

where

$$a = 2k^2 + 2M^2 + (\mu f)^2 + \left(\frac{\mu}{f}\right)^2 - 2(r_1 + r_2)\mu \frac{Pf}{1+f^2},$$

$$b = 2k\left(\left(\frac{\mu}{f}\right)^2 - (\mu f)^2 + 2(r_1 - r_2)\mu \frac{Pf}{1+f^2}\right),$$

$$c = k^4 + \left(2M^2 - (\mu f)^2 - \left(\frac{\mu}{f}\right)^2 + 2(r_1 + r_2)\mu \frac{Pf}{1+f^2}\right)k^2 +$$

$$+ (\mu^2 - M^2)\left[(\mu^2 - M^2) + 4r_1 r_2 \left(\frac{Pf}{1+f^2}\right)^2 - 2\mu \frac{Pf}{1+f^2}\left(\frac{r_1}{f^2} + r_2 f^2\right)\right].$$

The found polynomial has four roots. If at least one of these roots possesses a nonzero imaginary part an exponential growth of the amplitudes with the perturbations appears. This means that particular solutions (3) and (4) are unstable. The four roots of the polynomial in $\nu$ can be written as

$$\nu_{1,2} = \frac{\sqrt{a+2s}}{2} \pm \sqrt{\frac{a-2s}{4} - \frac{b}{2\sqrt{a+2s}}},$$

$$\nu_{3,4} = -\frac{\sqrt{a+2s}}{2} \pm \sqrt{\frac{a-2s}{4} + \frac{b}{2\sqrt{a+2s}}},$$

where

$$s = \sqrt[3]{-\frac{g}{2} + \sqrt{\left(\frac{g}{2}\right)^2 - \left(\frac{d}{3}\right)^3}} + \sqrt[3]{-\frac{g}{2} - \sqrt{\left(\frac{g}{2}\right)^2 - \left(\frac{d}{3}\right)^3}} - \frac{a}{6},$$

$$g = \frac{a^3}{108} - \frac{ac}{3} - \frac{b^2}{8}, \quad d = \frac{a^2}{12} + c.$$

Now it's possible to investigate the instability of continuum waves (3) and (4) with analyzing a gain spectrum which can be obtained by equation [35]:

$$G = |\text{Im}(\nu)_{max}|, \qquad (10)$$

where $\text{Im}(\nu)_{max}$ denotes the imaginary part of the root with the largest imaginary part.

### 5. Gain spectrum

In this section MI of the waves in considered NOWA will be investigated with analyzing the gain spectrum (10). According to the structure of the array, three different cases of nonlinear response will be considered. There are (**i**) all waveguides of the array are nonlinear focusing or defocusing one, ($r_1 = r_2 > 0$ or $r_1 = r_2 < 0$); (**ii**) some kind of waveguides is focusing and another one – defocusing, ($r_1 = -r_2 \neq 0$); (**iii**) some kind of waveguides is linear and another one – nonlinear, ($r_1 = 0$, $r_2 \neq 0$ or $r_2 = 0$, $r_1 \neq 0$). Also, the influences of the power $P$ and the ratio of the backward to forward wave amplitude $f$ on the gain spectrum will be studied.

The instability of particular solutions (3) and (4) appears when the gain spectrum (10) takes nonzero real values. Thus, it's easy to investigate MI with analyzing the gain spectrum now. In figure 4 it is shown that solutions (3) and (4) with $q = \pi$, which corresponds to the edge of Brillouine zone, become unstable with increasing the power *P*. In this case the analytical expression for the gain spectrum has the following form:

$$G = \sqrt{4K^2 \left|\cos\frac{Q}{2}\right|\left(\theta_{NL} - \left|\cos\frac{Q}{2}\right|\right) - k^2} \quad , \quad \theta_{NL} = \frac{\sqrt{r_1 r_2}}{K} \frac{P|f|}{1+f^2} . \tag{11}$$

For the first Brillouine zone, the stability region lies into interval: $-2\arccos\theta_{NL} \leq Q \leq 2\arccos\theta_{NL}$; the instability regions lie into intervals: $-\pi < Q < -2\arccos\theta_{NL}$, $2\arccos\theta_{NL} < Q < \pi$ (figure 4a). One can find that the gain spectrum (11) is equal to zero at boundary points of these regions. The stability region vanishes when $\theta_{NL} > 1$, and particular solutions (3) and (4) become entirely unstable (under *k* = *0*) (figure 4b). From this it's easy to determine a threshold value of power *P*:

$$\theta_{NL} > 1 \Rightarrow \frac{\sqrt{r_1 r_2}}{K} \frac{P|f|}{1+f^2} > 1 \Rightarrow P_{Tr} = \frac{K}{\sqrt{r_1 r_2}} \frac{1+f^2}{|f|} . \tag{12}$$

If $P > P_{Tr}$, particular solutions (3) and (4) become unstable for the whole Brillouine zone. From expressions (11) and (12) it follows that the threshold value of power and stability/instability regions can be controlled by the ratio of the backward to forward wave amplitude *f*. For example, the threshold value of power can be very high if $|f| << 1$ ($|b| << |a|$) or $|f| >> 1$ ($|b| >> |a|$), therefore, the considered waves (3) and (4) may be stable at very high intensities of electromagnetic field in the waveguides. Minimum of the threshold value of power corresponds to $|f| = 1$ ($|b| = |a|$) and it is equal to $P_{Tr}^{\min} = 2K/\sqrt{r_1 r_2}$. Maximum of the nonlinear gain situates at points $Q_m = \pm 2\arccos(\theta_{NL}/2)$ under *k* = *0* and its maximum magnitude is equal to $G_m = G(Q_m) = K\theta_{NL}$ (figure 4a, b, c). If $\theta_{NL} \geq 2$ ($P \geq 2P_{Tr}$) the maximum of the nonlinear gain located at point $Q_m = 0$ where *k* = *0* and its value is equal to $G_m = 2K\sqrt{\theta_{NL} - 1}$ (figure 4d).

From expression (11) it follows that the case of defocusing waveguides is similar to the case of focusing waveguides. One can see that in the case of (**ii**) the solutions (3) and (4) with $q = \pi$ are unstable and in the case of (**iii**) they are stable, because $\theta_{NL} = 0$ and MI doesn't occur. The absence of instability in the last case is the unique feature of considered array of coupled waveguides. However, in NODC nothing of the kind was observed. NODC consists of two coupled waveguides. Nevertheless, when one of these waveguides is linear one and another waveguide is nonlinear, the MI process occurs [27]. The considered NOWA is a system of a large number of coupled NODCs (figure 1), thus, the periodicity in NOWA has lead to the absence of MI in comparison with NODC, when one kind of waveguides is linear and another one is nonlinear.

In figures 5 and 6 it's shown that the solutions (3) and (4) with $q = \pi/2$ and with $q = 0$ respectively, also become unstable with increasing the power *P*, when the all waveguides are focusing. The spectrum of considered waves (3) and (4) with $q = \pi/2$ and with $q = 0$ has the band gap (figure 3). There are no considerable changes in the gain spectrum in these cases (figure 5, 6), in comparison with previous case (figure 4), but the presence of the band gap leads to the increase of the threshold value of power $P$ (figure 5, 6). Thus, the modes with the band gap are stable in higher power levels unlike the modes without the band gap in the NOWA under consideration. Situation changes considerably when the all waveguides are defocusing. The presence of the band gap results to the rise of considerable stability area which doesn't change with increasing the power (figure 7). In particular, the waves with $q = 0$ has turned to be stable for the perturbations with $k=0$ (figure 7c, d). In the cases of (**ii**) and (**iii**) nothing notable happens. The instability takes place at high values of power (figure 8).

## 6. Conclusions

In this work waveguide array involving nonlinear waveguides with positive and negative refractive indices (figure 1) is considered. The modulation instability of the coupled forward and backward waves in this waveguide system is studied.

Coupled mode theory underlies the method of investigations. Particular solutions of the system of coupled mode equations are found. These solutions describe propagation of continuum waves in longitudinal direction of waveguide array. Linear spectrum of these waves has the band gap, but for values of transverse wave number corresponding to edges of Brillouine zone the spectrum has no gap (figure 3). These transverse wave numbers are corresponding to the Bragg resonance in one dimensional lattice.

Instability of the obtained particular solutions is investigated with the linear stability analysis. Properties of the modulation instability process are studied with analyzing the gain spectrum. Influence of different parameters: electromagnetic field power, ratio of the backward to forward wave amplitude and the nonlinearity parameters on the gain spectrum is investigated analytically and numerically. In case of Bragg resonance, when the spectrum of continuum waves has no band gap, these waves become unstable with increasing the power if all waveguides are nonlinear focusing or defocusing (figure 4). Threshold value of power (12), under which the continuum waves become entirely unstable, is obtained in this case. This value can be controlled with the ratio of backward to forward wave amplitude. When the ratio takes very small or very large values the threshold value of power becomes very high, thus the continuum waves can be stable at very high intensities of the electromagnetic field in waveguides. When one kind of waveguides is focusing and another one is defocusing, the continuum waves are unstable in case of the Bragg resonance. However, if the nonlinearity assists at one kind of waveguides only and another one is linear, these waves turn to be stable ones and modulation instability doesn't occur. The absence of instability in the last case is the distinctive feature of this array of coupled waveguides and is a consequence of periodicity. Really, in nonlinear oppositely-directional coupler nothing of the kind was observed [27]. When the spectrum of continuum waves has the band gap, they also become unstable with increasing the power if all waveguides are focusing (figure 5, 6), but the threshold value of power become higher than in the case of Bragg resonance. Thus, the modes with the band gap are stable in higher power levels unlike modes without the band gap in the waveguide array. In case of defocusing waveguides a rather considerable stability area appears in the gain spectrum and this area doesn't change with increasing the power (figure 7). This means that continuum waves are more stable in defocusing waveguides than in focusing when the spectrum has the band gap. When one kind of waveguides is focusing and another one is defocusing or when one kind of waveguides is linear and another one is nonlinear the continuum waves become unstable at high values of power (figure 8).

**Acknowledgments**

We are grateful to S. A. Doudchenko for useful and fruitful discussions. This work was supported by the Russian Scientists Found (project No. 14-22-00098).

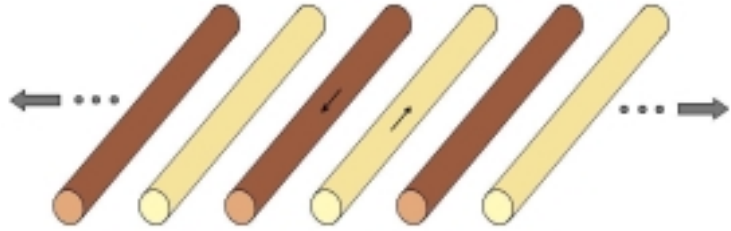

**Figure 1.** The waveguide array, (PIM waveguides are bright and NIM waveguides are dark).

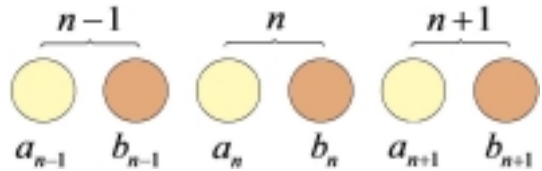

**Figure 2.** The waveguide array, (cross-section).

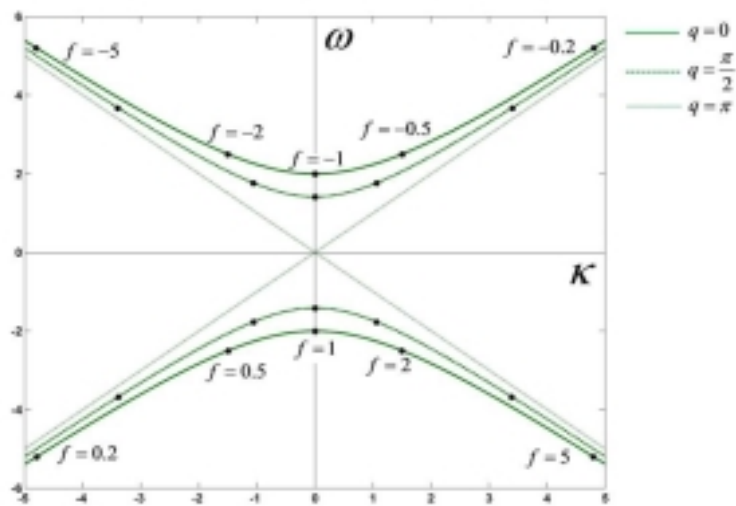

**Figure 3.** Dispersion relation of continuum waves propagating in the NOWA under fixed values of the transverse wave number $q$ ($\Delta=0$, $K=1$). Dots correspond to the nonlinear spectrum in the lower power limit: $r_1 = r_2 = 0.1$, $P = 0.1$.

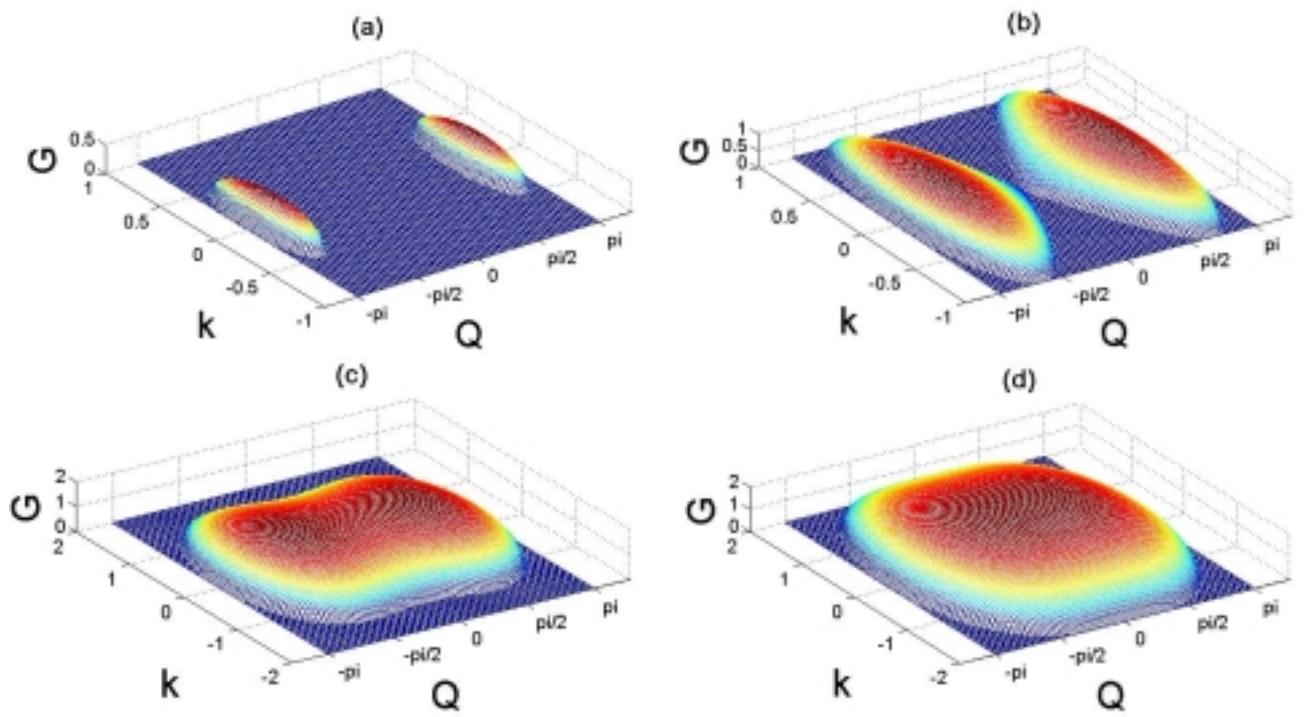

**Figure 4.** Gain spectrum vs transverse wave vector *Q* and propagation constant *k* of small perturbations under different values of power *P*, ($K = 1$, $r_1 = r_2 = 0.1$, $f = 1$, $q = \pi$). (a) $P = 10$, (b) $P = 20$, (c) $P = 30$, (d) $P = 40$.

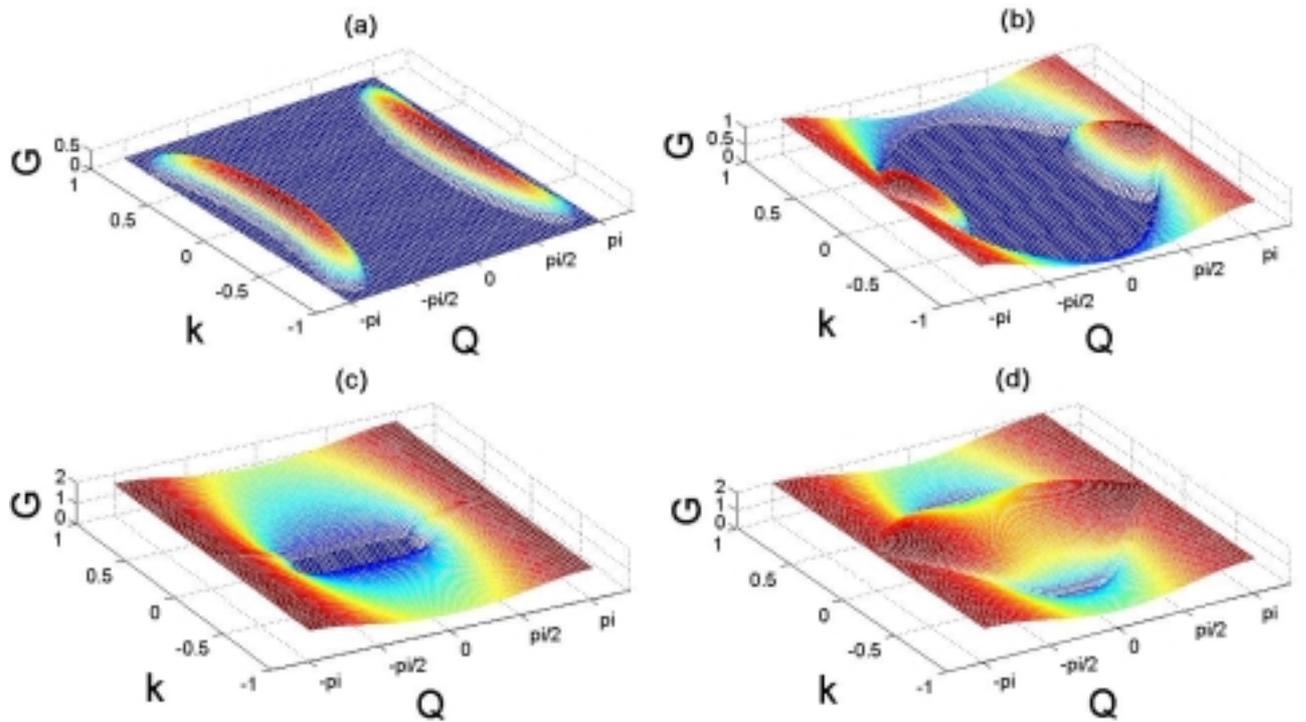

**Figure 5.** Gain spectrum vs transverse wave vector *Q* and propagation constant *k* of small perturbations under different values of power *P*, ($K = 1$, $r_1 = r_2 = 0.1$, $f = 1$, $q = \pi/2$). (a) $P = 10$, (b) $P = 20$, (c) $P = 30$, (d) $P = 40$.

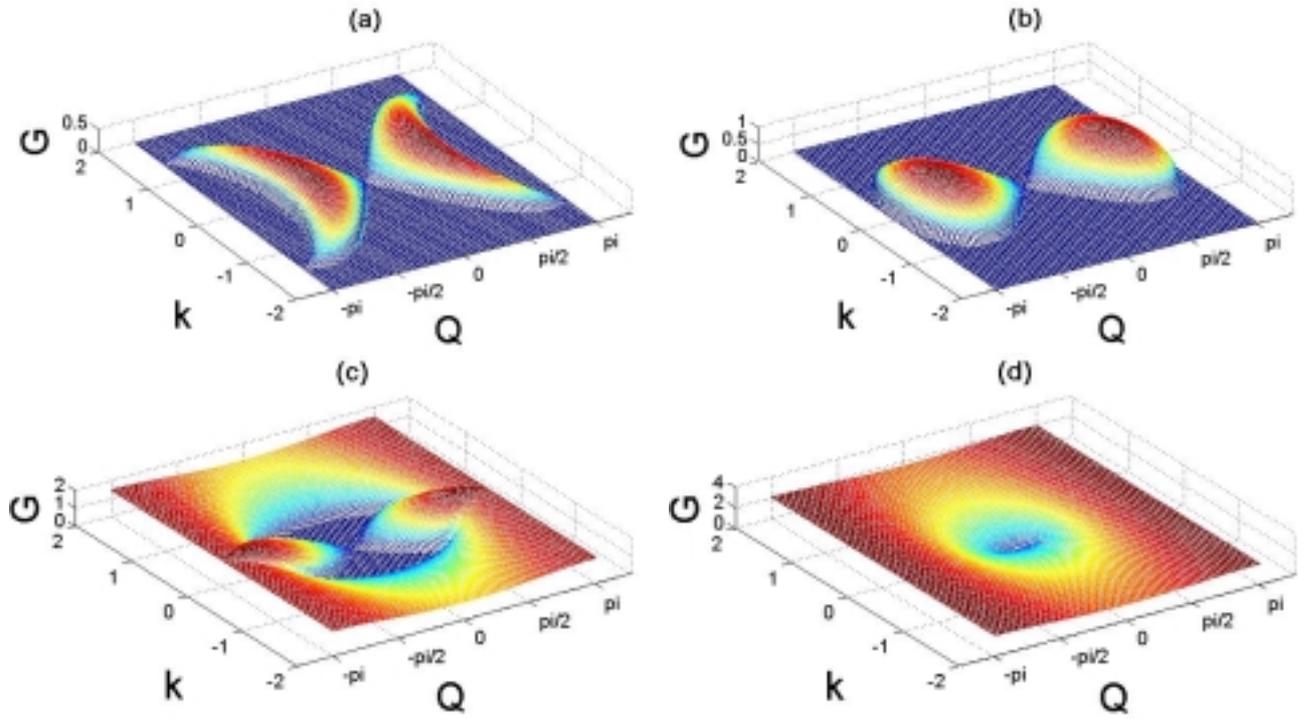

**Figure 6.** Gain spectrum vs transverse wave vector $Q$ and propagation constant $k$ of small perturbations under different values of power $P$, ($K = 1$, $r_1 = r_2 = 0.1$, $f = 1$, $q = 0$). (a) $P = 10$, (b) $P = 20$, (c) $P = 30$, (d) $P = 40$.

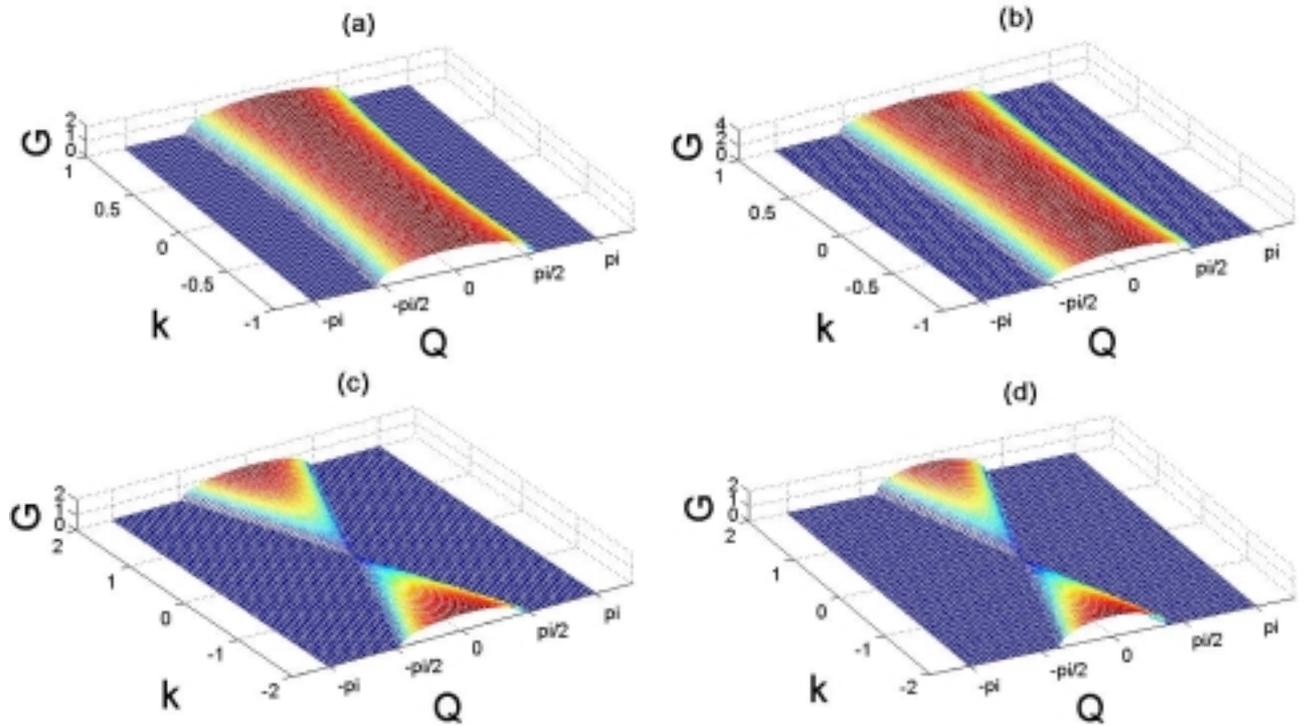

**Figure 7.** Gain spectrum vs transverse wave vector $Q$ and propagation constant $k$ of small perturbations under different values of power $P$ and of transverse wave number $q$, ($K = 1$, $r_1 = r_2 = -0.1$, $f = 1$). (a) $q = \pi/2$, $P = 40$, (b) $q = \pi/2$, $P = 80$, (c) $q = 0$, $P = 40$, (d) $q = 0$, $P = 80$.

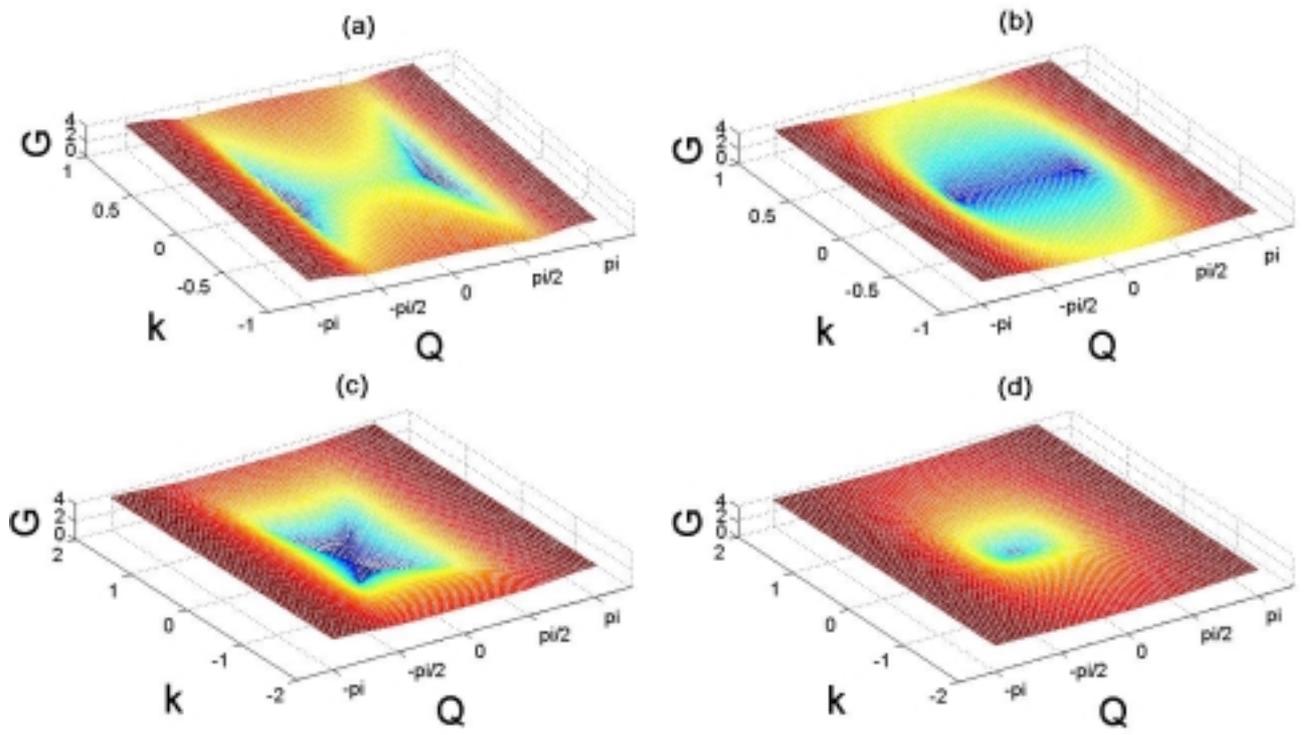

**Figure 8.** Gain spectrum vs transverse wave vector $Q$ and propagation constant $k$ of small perturbations under different nonlinear conditions and different values of transverse wave number $q$, ($K = 1, P=80, f = 1$). (a) $q = \pi/2$, $r_1 = -r_2=0.1$, (b) $q = \pi/2$, $r_1 = -r_2=0.1$, (c) $q = 0$, $r_1 =0.1$, $r_2=0$, (d) $q = 0$, $r_1 =0.1$, $r_2=0$.